\documentclass[runningheads]{llncs}
\usepackage{graphicx}
\usepackage{bm}
\usepackage{array}
\usepackage{amsmath}
\usepackage{float}
\usepackage[colorlinks=true,linkcolor=blue,citecolor=blue,urlcolor=blue,]{hyperref}
\usepackage[doipre={doi:~}]{uri}
\usepackage{framed,multirow}
\usepackage{booktabs}
\usepackage[perpage,symbol*]{footmisc}
\usepackage{marvosym}
\usepackage{float}
\begin{document}
\title{Learning Cross-Modal Deep Representations for Multi-Modal MR Image Segmentation}
\titlerunning{Cross-Modal Deep Representation Learning}
%
\author{Cheng Li\inst{*1} \and
Hui Sun\inst{*1} \and Zaiyi Liu\inst{2} \and Meiyun Wang\inst{3}
\and Hairong Zheng\inst{1} \and Shanshan Wang\inst{1}\textsuperscript{(\Letter)}}

\institute{Paul C. Lauterbur Research Center for Biomedical Imaging, Shenzhen Institutes of Advanced Technology, Chinese Academy of Sciences, Shenzhen, Guangdong, China\\
\email{sophiasswang@hotmail.com}\\
\and Department of Radiology, Guangdong General Hospital, Guangdong Academy of Medical Sciences, Guangzhou, Guangdong, China\\
\and Department of Medical Imaging, Henan Provincial People’s Hospital, Zhengzhou, Henan, China\\}

\authorrunning{C. Li et al.}

\maketitle              

\renewcommand{\thefootnote}{}
\footnotetext{* These authors contributed equally to this work.} 

\begin{abstract}
Multi-modal magnetic resonance imaging (MRI) is essential in clinics for comprehensive diagnosis and surgical planning. Nevertheless, the segmentation of multi-modal MR images tends to be time-consuming and challenging. Convolutional neural network (CNN)-based multi-modal MR image analysis commonly proceeds with multiple down-sampling streams fused at one or several layers. Although inspiring performance has been achieved, the feature fusion is usually conducted through simple summation or concatenation without optimization. In this work, we propose a supervised image fusion method to selectively fuse the useful information from different modalities and suppress the respective noise signals. Specifically, an attention block is introduced as guidance for the information selection. From the different modalities, one modality that contributes most to the results is selected as the master modality, which supervises the information selection of the other assistant modalities. The effectiveness of the proposed method is confirmed through breast mass segmentation in MR images of two modalities and better segmentation results are achieved compared to the state-of-the-art methods.

\keywords{Supervised feature fusion \and Multi-modal image segmentation \and Spatial attention.}
\end{abstract}

\section{Introduction}
Multi-modal magnetic resonance imaging (MRI) is an essential tool in clinics for the screening and diagnosis of different diseases including breast cancer, prostate cancer, and neurodegenerative disorders. The combination of different imaging modalities can overcome the limitations of the individual modalities. In breast cancer screening, for example, while contrast-enhanced MRI possesses high sensitivity in detecting breast lesions, T2-weighted MRI is effective in reducing false-positive results \cite{ref_article1,ref_article2}. Considering different MRI modalities is thus important for the acquisition of accurate lesion information. Lesion segmentation of MR images is a critical step in the process for the following diagnosis and surgical planning. Manual segmentation is both time-consuming and error-prone. Therefore, the development of automatic and reliable algorithms is of high clinical values.

Learning-based methods, especially those based on convolutional neural networks (CNNs), have seen rapid development in medical image analysis in the last decade \cite{ref_article3}. CNNs were originally proposed for the task of image-level classification. The intuitive application of CNNs to image segmentation, which is a pixel-level classification task, was conducted by classifying each pixel in a sliding window manner (R-CNN) \cite{ref_article4}. Fully convolutional neural networks (FCNs) were designed later to avoid the cumbersome and memory-deficient R-CNN approach \cite{ref_article5}. FCNs segment the input image directly by generating heatmap output. Following FCNs, U-Net was proposed specifically for biomedical image segmentation \cite{ref_article6}, which is the current baseline network for various medical image segmentation tasks and is the inspiration of many subsequent works.

A critical issue regarding multi-modal image segmentation is the fusion of information from the different imaging modalities. CNN-based multi-modal image fusion can be realized through early fusion, late fusion, and multi-layer fusion. Early fusion happens at the input stage or low-level feature stages \cite{ref_article7,ref_article8}. This strategy may fail to achieve the expected information compensation, especially when the different modal images have complex relationships. Late fusion refers to the fusion of high-level and high-abstract features, and multi-stream networks are commonly utilized in this case with each stream processing images from one modality. Late fusion has been demonstrated to generate better segmentation results than direct early fusion \cite{ref_article9,ref_article10}. Nevertheless, multi-layer fusion should be a more generalized strategy. Multi-layer fusion was first proposed for RGB-D image segmentation where FuseNet was designed to incorporate depth information into RGB images \cite{ref_article11}. Further network optimization over FuseNet confirmed that multi-layer fusion was a more effective approach \cite{ref_article12}. Multi-layer fusion has also been successfully applied to multi-modal medical image segmentation \cite{ref_article13}. Although inspiring results have been achieved, the feature fusion was conducted through direct pixel-wise summation or channel-wise concatenation. Without supervision and selection, the fusion process may introduce irrelevant signals and noise signals to the final outputs.

In this study, we propose a novel multi-stream CNN-based feature fusion network for the processing of multi-modal MR images. In accordance with real clinical situations, we pick one MR modality that contributes most to the final segmentation results as the master modality, and the other modalities are treated as assistant modalities. Inspired by the knowledge distillation concept \cite{ref_article14}, where a teacher network supervises the training of a student network, our master modal network stream supervises the training of the assistant modal network streams. In detail, we adopt an attention block to extract the supervision information from the master modality and utilize this supervision information to select useful information from both the master modality and the assistant modalities. The effectiveness of the proposed method is evaluated through the mass segmentation in breast MR images of two modalities. Segmentation of breast mass structures is a challenging task, as the masses have a large range of sizes and shapes, especially for spiculated masses that have ill-defined borders. The results show that our method can achieve the best performance compared to existing feature fusion strategies.

\section{Methodology}
\subsubsection{Breast MRI Dataset.} The breast MR images were collected using an Achieva 1.5T system (Philips Healthcare, Best, Netherlands) with a four-channel phased-array breast coil. All acquisitions of 313 patients in the prone position were conducted between 2011 and 2017. Two MRI sequences were applied. Axial T2-weighted (T2W) images (TR/TE = 3400 ms/90 ms, FOV = 336 mm $\times$ 336 mm, section thickness = 1 mm) with fat suppression were obtained before the injection of contrast medium. After the intravenous injection of 0.3 mL/kg of gadodiamide (BeiLu Pharmaceutical, Beijing, China), axial fat-suppressed contrast-enhanced T1-weighted (T1C) images were collected (TR/TE = 5.2 ms/2.3 ms, FOV = 336 mm $\times$ 336 mm, section thickness = 1 mm, and flip angle = $15^o$). Since manual segmentation of the breast masses in 3D multi-modal MR images is very difficult and time-consuming, only the central slices with the largest cross-section areas were labelled by two experienced radiologists in this retrospective study.

\begin{figure}[t]
\includegraphics[width=\textwidth]{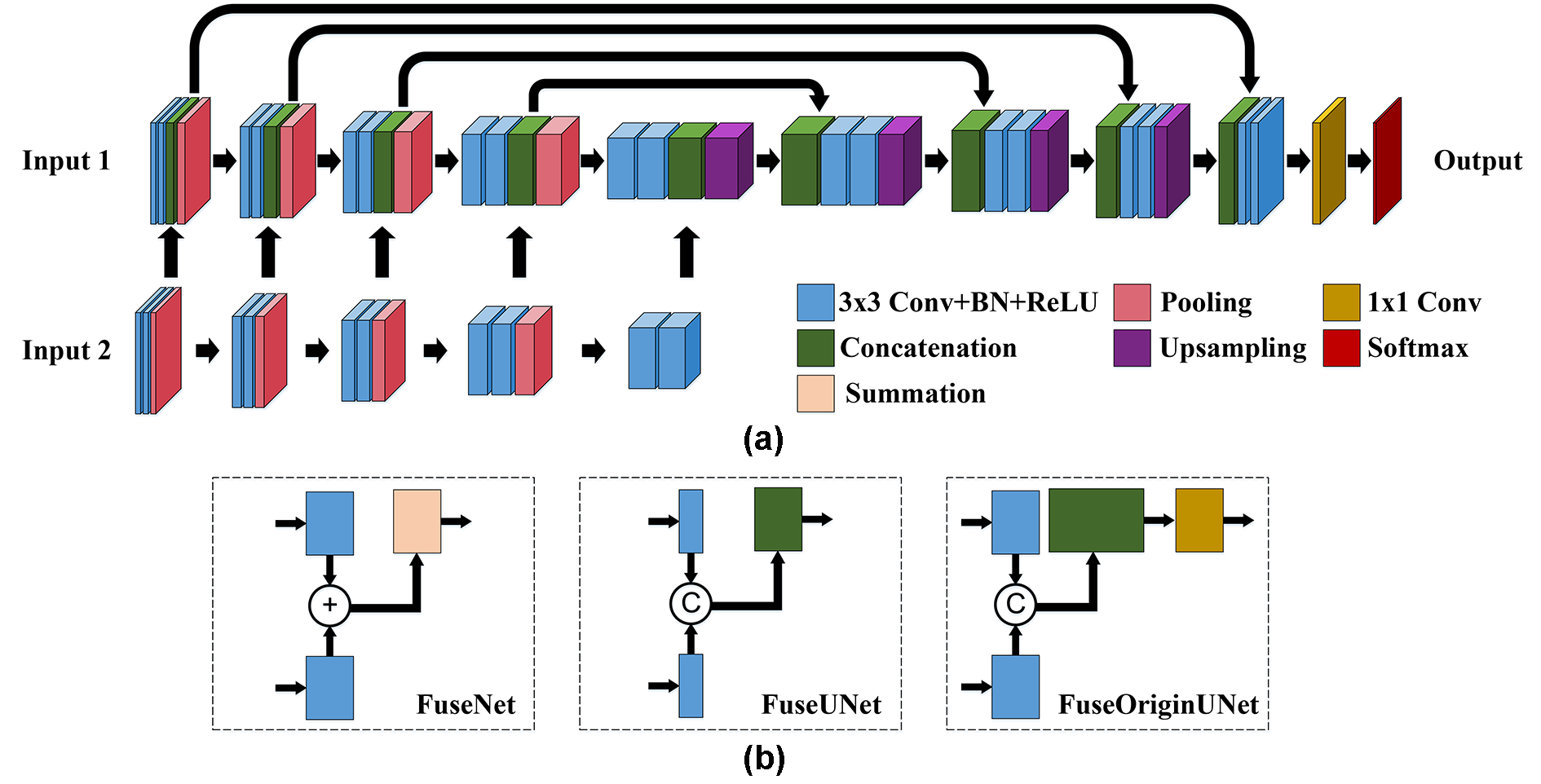}
\caption{Baseline network architecture.  The overall network structure (a) and the implementation details of the modality fusion by the three networks (b).} \label{fig1}
\end{figure}

\subsubsection{A Better Feature Fusion Baseline Network Architecture.} Our baseline model is built from FuseNet \cite{ref_article11} with two major modifications. First, FuseNet was proposed for the analysis of natural images. The encoder part of FuseNet adopted the VGG 16-layer model for the convenience of utilizing ImageNet pre-trained network parameters. To better adapt to medical image processing, we build a FuseNet-like network architecture based on U-Net, named FuseOriginUNet (Fig.~\ref{fig1}). Second, in FuseNet, the feature fusion of different imaging modalities was realized by pixel-wise summation, which could preserve the VGG 16-layer model after introducing the feature fusion module. In FuseOriginUNet, a channel-wise concatenation is implemented instead. To make the overall network lightweight, we half the convolution kernels for each layer in the encoder part compared to U-Net and achieve the final baseline model FuseUNet (Fig.~\ref{fig1}). The experiments show that FuseUNet achieves better performance compared to both FuseNet and FuseOriginUNet.

\begin{figure}[t]
\includegraphics[width=\textwidth]{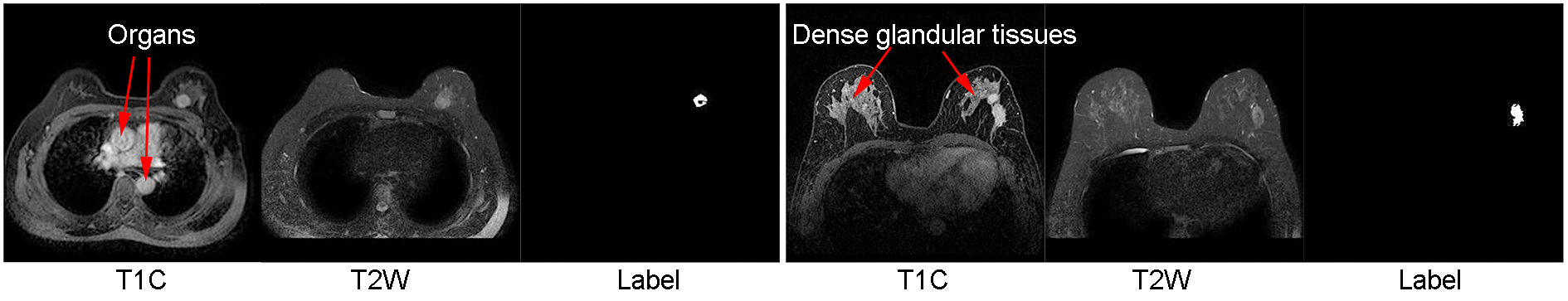}
\caption{Example breast MR images when T1C images highlight irrelevant regions (organs or dense glandular tissues) and T2W can distinguish these regions from the targeted breast masses. Label images are the manual segmentation results.} \label{fig2}
\end{figure}

\subsubsection{Supervised Cross-Modal Deep Representation Learning.} Different imaging modalities contain different sorts of useful information for the targeted task. For breast MR images, the T1C modality has a high sensitivity and a relatively low specificity in detecting breast masses. Two examples are shown in Fig.~\ref{fig2}. It can be observed that the T1C image highlights not only the breast mass area but also the irrelevant regions, such as the organs and the dense glandular tissues. In this case, T2W images are important in distinguishing the true masses from all the enhanced areas. Accordingly, the two imaging modalities are treated differently in the proposed method. T1C is chosen as the master modality having a greater impact on the results. T2W is regarded as the assistant modality complementing the information of the master modality.

\begin{figure}
\includegraphics[width=\textwidth]{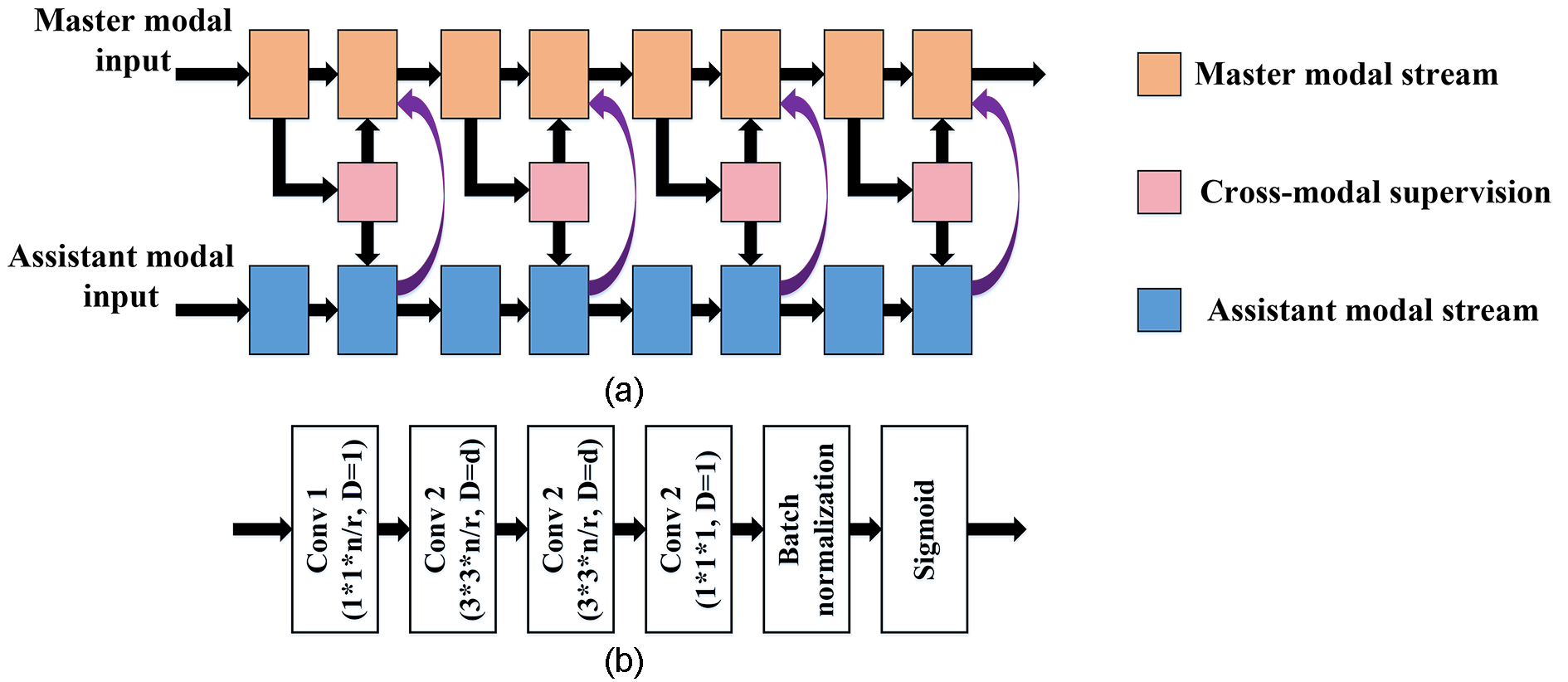}
\caption{The encoder section of the proposed master--assistant cross-modal learning network (a) and the cross-modal supervision learning module (n is the input feature number, r is a reduction factor, and D is the atrous rate of dilated convolutions) (b).} \label{fig3}
\end{figure}

Inspired by the knowledge distillation between teacher--student networks \cite{ref_article14}, we propose a supervised master--assistant cross-modal learning framework (Fig.~\ref{fig3}a). The master modality generates supervision information that modulates the learning of the assistant modality. Enlightened by the activation-based attention transfer strategies \cite{ref_article15}, a spatial attention (SA) block is designed to extract the supervision information (Fig.~\ref{fig3}b). The input of the block is the features from the master modal stream and the output, which is a weight heatmap, is utilized to guide the information selection for both the master and the assistant modalities.

\subsubsection{Implementation Details.} Five-fold cross-validation experiments were conducted. All the images along with the label images were resized to $256 \times 256$ and intensity normalized. No further data processing or augmentation was applied. The models were implemented with PyTorch on a NVIDIA TITAN Xp GPU (12G) with batch size of 4. ADAM with AMSGRAD was applied to train the models. The step decay learning rate strategy was used with an initial learning rate of 1e-4 that was decreased by half every 30 epochs. The hyperparameters in the SA module were set as r=16 and d=4. To tackle the widely recognized class-imbalance problem in medical image analysis, a loss function combining cross-entropy loss and Dice loss was adopted:
\noindent
\begin{equation}
\begin{aligned}
L & =L_{Dice}+\alpha \times L_{CE} \\
& =(1-\frac{2\sum_{i=1}^Np_iy_i+\varepsilon}{\sum_{i=1}^Np_i+\sum_{i=1}^Ny_i+\varepsilon}) + \alpha \times (-\frac{1}{N}(y_i\sum_{i=1}^Np_i+(1-y_i)\sum_{i=1}^N(1-p_i)))
\end{aligned}
\end{equation}
where $L$ is the loss function utilized, $L_{Dice}$ is the Dice loss, $L_{CE}$ is the cross-entropy loss, $N$ is the total number of pixels in the image, $y_i\in \{0,1\}$ is the manual segmentation label of the $i^{th}$ pixel in the image where 0 refers to the background and 1 refers to the foreground, $p_i\in [0,1]$ is the corresponding predicted probability of the $i^{th}$ pixel belonging to the foreground class, $\varepsilon = 1.0$ is a constant to keep the numerical stability, and $\alpha = 1.0$ is a weight constant to control the tradeoff between the two losses.

Three metrics were utilized to quantify the segmentation performance, the Dice similarity coefficient, sensitivity, and relative area difference. Three independent experiments were run, and the results are presented as ($mean \pm s.d.$).

\begin{table}
\caption{Segmentation results of different models.}\label{tab1}
\begin{tabular}{c|p{2.2cm}<{\centering}|p{2.2cm}<{\centering}|p{2.2cm}<{\centering}|p{2.2cm}<{\centering}}
\hline
Models & Number of parameters$^*$ & Dice$^\#$ & Sensitivity$^\#$ & Relative area difference$^\#$ \\
\hline
U-Net (T1C) & 34.5M  & $73.3 \pm 0.1$ & $79.5 \pm 0.8$ & $43.7 \pm 1.9$\\
U-Net + SA (T1C) & 34.7M & $75.8 \pm 0.2$ & $82.6 \pm 0.5$ & $33.9 \pm 2.2$\\
\hline
FuseNet & 29.4M & $52.8 \pm 1.5$ & $54.9 \pm 2.7$ & $49.2 \pm 2.7$\\
FuseNetConcate & 74.0M & $73.2 \pm 0.3$ & $80.9 \pm 0.8$ & $41.8 \pm 1.3$\\
FuseOriginUNet & 56.2M & $75.0 \pm 0.5$ & $82.1 \pm 0.5$ & $38.0 \pm 4.4$\\
EarlyFuseUNet & 34.5M & $74.4 \pm 0.2$ & $81.4 \pm 1.0$ & $38.1 \pm 4.8$\\
LateFuseUNet & \textbf{25.1M} & $73.8 \pm 0.1$ & $81.4 \pm 1.7$ & $39.3 \pm 4.1$\\
FuseUNet & 26.7M & $74.9 \pm 0.2$ & $82.2 \pm 0.4$ & $37.9 \pm 2.5$\\
FuseUNet + SA & 26.8M &  $ 76.3 \pm 0.2$ & $83.4 \pm 0.4$ & $32.4 \pm 0.8$\\
\hline
{\bfseries Proposed }& 26.7M & $\bm{77.6 \pm 0.3}$ & $\bm{84.4 \pm 0.7}$ & $\bm{30.9 \pm 1.6}$\\
\hline
\multicolumn{5}{@{}l}{$*$ M--millions. $\#$ Values in percentage.} \\
\end{tabular}
\end{table}

\section{Results and Discussion}

Table~\ref{tab1} lists the quantitative segmentation results of the different networks. It can be concluded that compared to the pixel-wise summation strategy used in FuseNet, multi-modal feature fusion by channel-wise concatenation (FuseNetConcate) is more effective. Adopting the U-Net blocks (FuseOriginUNet) leads to further performance enhancement. Moreover, the lightweight FuseUNet achieves a comparable or even superior level of segmentation accuracy with only half of the parameters used by FuseOriginUNet. 
U-Net trained solely on T1C presents worse performance than all the two-modal U-Net based networks, suggesting that T2W images 
\begin{figure}
\includegraphics[width=\textwidth]{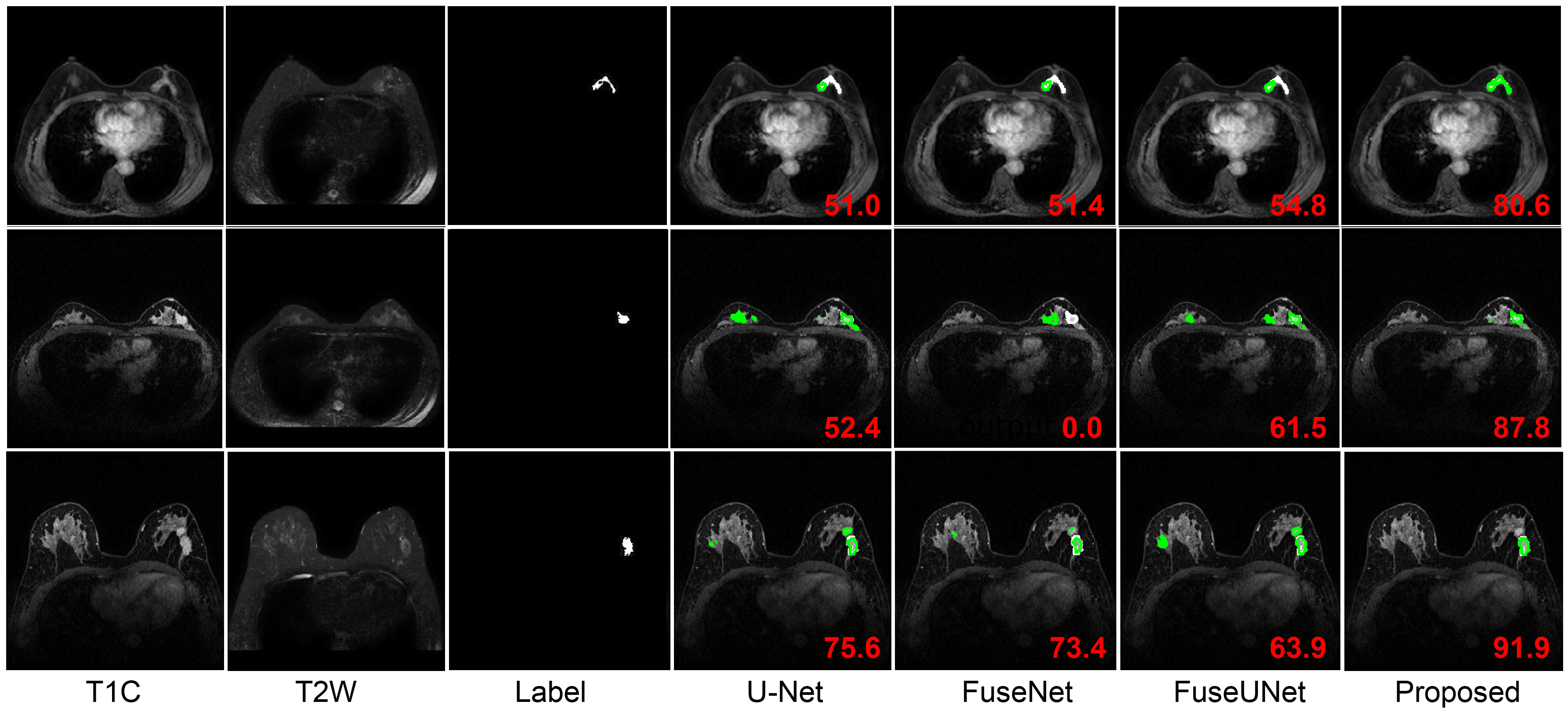}
\caption{Example results of the different networks. White lines indicate the boundaries of the manual segmentation labels. Green lines are the boundaries of the network segmentation results. The value in each image is the Dice similarity coefficient (\%).} \label{fig4}
\end{figure}
provided useful complementary information for the segmentation task. For the two-modal U-Net based networks, multilayer fusion of FuseNet is more effective than both early fusion (EarlyFuseUNet) and late fusion (LateFuseUNet). Introducing our SA block to both U-Net (U-Net + SA) and FuseUNet (FuseUNet + SA) before each pooling layer elevates segmentation performance. Finally, our proposed supervised cross-modal deep representation learning method generates the best segmentation results reflected by all three metrics.

The segmentation results of several examples are given in Fig.~\ref{fig4}. Overall, models utilizing two modal inputs are more effective than the single-modal U-Net. Except in the last example, the improved baseline model FuseUNet achieves a higher Dice similarity coefficient than FuseNet. The proposed method consistently achieves much better results than the existing methods with decreased false negatives (first example) and decreased false positives (second and third examples).

\begin{figure}
\includegraphics[width=\textwidth]{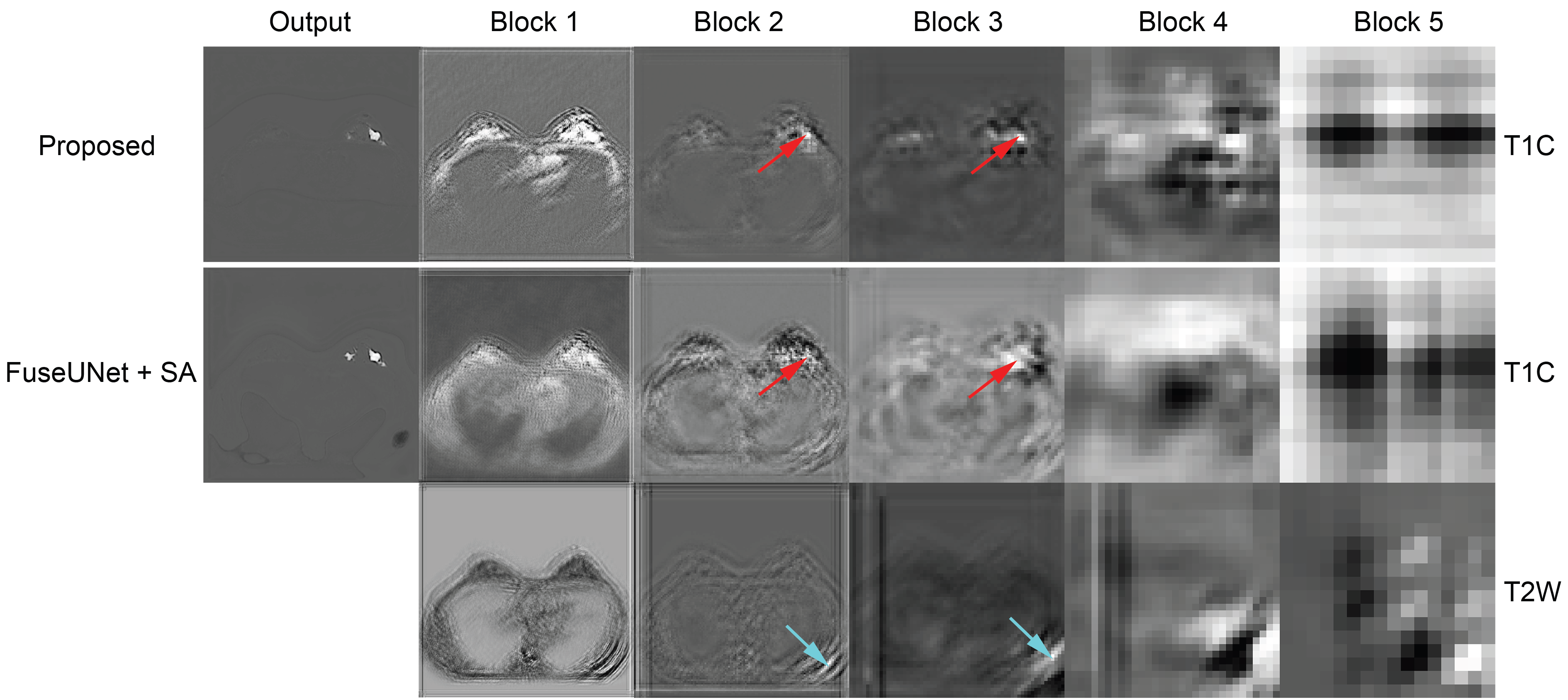}
\caption{SA maps of the proposed method and the FuseUNet + SA model. Blocks 1-5 refer to the attention maps generated at the five blocks before the respective pooling layers. T1C and T2W refer to the feature maps generated by the T1C modal stream and the T2W modal stream.} \label{fig5}
\end{figure}

To demonstrate the mechanism behind the improved performance brought by the proposed method, the SA maps of all five down-sampling blocks are visualized. One example is presented in Fig.~\ref{fig5}. It is clear from Fig. 5 that the T1C modal stream in both the proposed method and the FuseUNet + SA model was able to localize the mass regions through implementing the SA modules (red arrows in Fig.~\ref{fig5}). The T2W modal stream could hardly find the interesting areas and even highlighted the regions that were irrelevant for the task (blue arrows in Fig.~\ref{fig5}). Therefore, it is reasonable and necessary to apply the T1C attention maps to the information selection of T2W. For situations where different modalities generate images with similar sensitivites, our network architecture can still be utilized with an accordinly designed supervision information extraction strategy. The main idea regarding the supervised feature fusion of different imaging modalities should always be beneficial.

\section{Conclusion}
In this work, we presented a novel network for the segmentation of multi-modal MR images. Inspired by the knowledge distillation and attention transfer strategies, a supervised cross-modal deep representation learning method was designed that selectively fused the useful information from the different modalities and suppressed the respective noise signals. Results on an in-vivo breast MR image dataset of two modalities confirmed the effectiveness of the proposed method.The proposed method is extendable to different medical image segmentation scenarios and will be investigated in the future.

\subsubsection{Acknowledgements.} 
This research was partially supported by the National Natural Science Foundation of China (61601450, 61871371, 81830056), Science and Technology Planning Project of Guangdong Province (2017B020227012, 2018B010109009), Youth Innovation Promotion Association Program of Chinese Academy of Sciences (2019351), and the Basic Research Program of Shenzhen (JCYJ20180507182400762).

%
%
%

\end{document}